\documentstyle[twoside]{article}

\catcode`\@=11
\long\def\@makefntext#1{
\protect\noindent \hbox to 3.2pt {\hskip-.9pt  
$^{{\eightrm\@thefnmark}}$\hfil}#1\hfill}		

\def\@makefnmark{\hbox to 0pt{$^{\@thefnmark}$\hss}}	
	
\def\ps@myheadings{\let\@mkboth\@gobbletwo
\def\@oddhead{\hbox{}
\rightmark\hfil\eightrm\thepage}   
\def\@oddfoot{}\def\@evenhead{\eightrm\thepage\hfil
\leftmark\hbox{}}\def\@evenfoot{}
\def\sectionmark##1{}\def\subsectionmark##1{}}

\oddsidemargin=\evensidemargin
\newcounter{sectionc}\newcounter{subsectionc}\newcounter{subsubsectionc}
\renewcommand{\section}[1] {\vspace{12pt}\addtocounter{sectionc}{1} 
\setcounter{subsectionc}{0}\setcounter{subsubsectionc}{0}\noindent 
	{\tenbf\thesectionc. #1}\par\vspace{5pt}}
\renewcommand{\subsection}[1] {\vspace{12pt}\addtocounter{subsectionc}{1} 
	\setcounter{subsubsectionc}{0}\noindent 
	{\bf\thesectionc.\thesubsectionc. {\kern1pt \bfit #1}}\par\vspace{5pt}}
\renewcommand{\subsubsection}[1] {\vspace{12pt}\addtocounter{subsubsectionc}{1}
	\noindent{\tenrm\thesectionc.\thesubsectionc.\thesubsubsectionc.
	{\kern1pt \tenit #1}}\par\vspace{5pt}}
\newcommand{\nonumsection}[1] {\vspace{12pt}\noindent{\tenbf #1}
	\par\vspace{5pt}}

\newcommand{\textlineskip}{\baselineskip=13pt}
\newcommand{\smalllineskip}{\baselineskip=10pt}

\def\eightcirc{
\begin{picture}(0,0)
\put(4.4,1.8){\circle{6.5}}
\end{picture}}
\def\eightcopyright{\eightcirc\kern2.7pt\hbox{\eightrm c}} 

\newcommand{\copyrightheading}[1]
	{\vspace*{-2.5cm}\smalllineskip{\flushleft
        {\footnotesize Los Alamos archives: gr-qc/9405072, revised and updated
        #1}\\
	 }}

\def\abstracts#1#2#3{{
	\centering{\begin{minipage}{4.5in}\baselineskip=10pt\footnotesize
	\parindent=0pt #1\par 
	\parindent=15pt #2\par
	\parindent=15pt #3
	\end{minipage}}\par}} 

\renewenvironment{thebibliography}[1]
	{\frenchspacing
	 \ninerm\baselineskip=11pt
	 \begin{list}{\arabic{enumi}.}
        {\usecounter{enumi}\setlength{\parsep}{0pt}     
	 \setlength{\leftmargin 12.7pt}{\rightmargin 0pt} 
         \setlength{\itemsep}{0pt} \settowidth
	{\labelwidth}{#1.}\sloppy}}{\end{list}}

\newcounter{itemlistc}
\newcounter{romanlistc}
\newcounter{alphlistc}
\newcounter{arabiclistc}

\def\@citex[#1]#2{\if@filesw\immediate\write\@auxout
	{\string\citation{#2}}\fi
\def\@citea{}\@cite{\@for\@citeb:=#2\do
	{\@citea\def\@citea{,}\@ifundefined
	{b@\@citeb}{{\bf ?}\@warning
	{Citation `\@citeb' on page \thepage \space undefined}}
	{\csname b@\@citeb\endcsname}}}{#1}}

\newif\if@cghi
\def\cite{\@cghitrue\@ifnextchar [{\@tempswatrue
	\@citex}{\@tempswafalse\@citex[]}}
\def\citelow{\@cghifalse\@ifnextchar [{\@tempswatrue
	\@citex}{\@tempswafalse\@citex[]}}
\def\@cite#1#2{{$\null^{#1}$\if@tempswa\typeout
	{IJCGA warning: optional citation argument 
	ignored: `#2'} \fi}}

\def\@refcitex[#1]#2{\if@filesw\immediate\write\@auxout
	{\string\citation{#2}}\fi
\def\@citea{}\@refcite{\@for\@citeb:=#2\do
	{\@citea\def\@citea{, }\@ifundefined
	{b@\@citeb}{{\bf ?}\@warning
	{Citation `\@citeb' on page \thepage \space undefined}}
	\hbox{\csname b@\@citeb\endcsname}}}{#1}}

\def\@refcite#1#2{{#1\if@tempswa\typeout
        {IJCGA warning: optional citation argument
	ignored: `#2'} \fi}}

\def\refcite{\@ifnextchar[{\@tempswatrue
	\@refcitex}{\@tempswafalse\@refcitex[]}}

\def\pmb#1{\setbox0=\hbox{#1}
	\kern-.025em\copy0\kern-\wd0
	\kern.05em\copy0\kern-\wd0
	\kern-.025em\raise.0433em\box0}

\def\fnt#1#2{\footnotetext{\kern-.3em
	{$^{\mbox{\scriptsize #1}}$}{#2}}}

\font\tenrm=cmr10
\font\tenit=cmti10 
\font\tenbf=cmbx10
\font\bfit=cmbxti10 at 10pt
\font\ninerm=cmr9

\font\eightrm=cmr8

\def\qed{\hbox{${\vcenter{\vbox{			
   \hrule height 0.4pt\hbox{\vrule width 0.4pt height 6pt
   \kern5pt\vrule width 0.4pt}\hrule height 0.4pt}}}$}}

\begin{document}



\normalsize\textlineskip
\thispagestyle{empty}
\setcounter{page}{1}

\copyrightheading{}                     

\vspace*{0.88truein}

\centerline{\bf BLACK HOLE THERMAL EFFECTS IN THE SCATTERING PICTURE}
\vspace*{0.035truein}
\vspace*{0.37truein}
\centerline{\footnotesize HARET C. ROSU}
\vspace*{0.015truein}
\centerline{\footnotesize\it Instituto de F\'{\i}sica,
Universidad de Guanajuato, Apdo Postal E-143, Le\'on, Gto, Mexico}
\baselineskip=10pt
\vspace*{10pt}
\vspace*{0.225truein}

\vspace*{0.21truein}
\abstracts{
In the scattering picture, thermal radiation effects may be associated with
the `above barrier' reflection coefficient, which comes into play because
of complex turning points. This note contains several general remarks on the
application of the above statement to Schwarzschild black hole radiance.\\
}{}{}


\textlineskip                  
\vspace*{12pt}                 

\vspace*{1pt}\textlineskip	
\vspace*{-0.5pt}
\noindent


\noindent




\noindent






{\bf 1}.- At the present time,
Hawking effect,\cite{h} and Unruh effect,\cite{u} are well settled paradigms
in quantum field theory and in physics in general.\cite{t} The widespread
opinion is that they are a significant contribution to our knowledge of the
important topic of {\em quantum field thermality}.

In this note, I comment on the occurence of
thermal effects in the scattering picture, in particular their connection
with the `above-barrier' reflection coefficient.
This seems to be less known than, for example, the fact
that superradiant scattering is closely analogous to stimulated emission.


{\bf 2}.-
To derive the Schwarzschild `thermal' radiance Hawking used a semiclassical
splitting of the massless Klein-Gordon solution together with an
asymptotic, {\em i.e.}, {\em in}-{\em out} formalism, which in fact
is quite alien to the WKB approach.
Following a discussion by Schoutens et al,\cite{svv} one can say that
Hawking discovered
that for massless scalar $s$-waves propagating on the Schwarzschild
background a broad {\em in}-signal at past null
infinity $\cal I^{-}$ of the form,
$$
\exp (i\omega v(u))=\exp (i\omega u) [(u_{0}-u)/4M]^{-i4M\omega}
\theta (u_{0}-u)  \eqno(1)
$$
ends up into an outgoing $s$-wave with a given frequency $\omega$ at future
null infinity $\cal I^{+}$.
In (1), $v=t+r^{*}$ and $u=t-r^{*}$ are null cone coordinates
and $r^{*}=(2M)[\frac{r}{2M}+\ln|\frac{r}{2M}-1|]$ is the tortoise coordinate.
The broad {\em in}-signal decomposes into a linear
superposition of `plane' waves with very different frequencies, and this implies
non-trivial Bogolubov transformations between the Fourier coefficients of
the {\em out} and {\em in} fields. The physical result is just the Hawking
`thermal' radiation.

{\bf 3}.- Consider now the case of $s$-wave scattering in Schwarzschild
geometry. The massless scalar equation in Schwarzschild
metric for $s$-waves of {\em fixed} frequency $\omega$ can be written as a
one-dimensional Schr\"odinger equation of the form,\cite{m} (with $c=\hbar =1$,
and choosing length dimensions for the radial and time coordinates as well as
for the black hole `mass')
$$
\Big[\frac{d^{2}}{dr^{*2}}+ \omega ^{2} -V_{s}(r^{*})\Big]\psi=0
\eqno(2)
$$
where $V_{s}=(1-\frac{2M}{r})\frac{2M}{r^{3}}$.
The potential $V_{s}$ is of finite range, in the sense that it
goes to
zero for both $\infty$ and $-\infty$ in the $r^{*}$ coordinate, and has been
plotted by Matzner,\cite{m} in 1968. It has a
low peak $V_{max} =3(\frac{3}{16})^{2} (2M)^{-2}$ at $r^{*}=0.23(2M)$,
(the horizon is to be found at $r^{*}\rightarrow -\infty$).
As I said, the main point is that one may associate thermal
radiation effects (Hawking radiance)
with the `above barrier' reflection coefficient and if Matzner would have
considered this case he would have discovered the thermal radiation effect in
1968.
There are very few authors who previously provided arguments related
to the `above barrier' coefficient.
Stephens,\cite{s} used an inverted oscillator barrier (for which
all $\omega > 0$ correspond to `above barrier' scattering),
while Grishchuk and Sidorov,\cite{gs} employed the squeezing approach.
Stephens' argument is based on the formal change of an evolution operator
in a density matrix if one replaces $it/\hbar$ by
$\beta$. This transformation is the result of the substitutions
$t\rightarrow -i\tau$, $x\rightarrow x$, $\dot x\rightarrow -i\dot x$,
$V(x)\rightarrow -V(x)$, and $E\rightarrow -E$, leading to the replacement of
the barrier by its upside-down counterpart.
As for the squeezing approach,
Hawking's squeezing effect in the quasistationary region outside
the horizon refers to the so-called {\em w} and {\em y} quasi-particles (those
having zero Cauchy data on past null infinity, and zero Cauchy data on past
horizon for positive, respectively negative Kruskal parameter). Moreover, one
way of writing a black hole {\em in}-vacuum state in terms of the
{\em out}-vacuum state is by means of a
squeezing operator
of the type $S(r,\pi)|0_{w}\rangle|0_{y}\rangle$, where the EPR-correlations are
manifest in the phase of $\pi$ of the squeeze operator.
The Hawking-Bogolubov coefficient,\cite{gs,h2}
can be identified with the `above-barrier' reflection coefficient
$$
x_{hb}=\exp(-8\pi GM\omega)=\exp(-\omega/T_{h})=R_{\omega}~.
\eqno(3)
$$
It is known that for energies well above the potential
barrier the semiclassical reflection coefficient becomes exponentially small.
Thus, the slope of the `above barrier' reflection coefficient in a logarithmic
scale is the same
for all high frequencies and may be used to define a  unique temperature
parameter.
Only in this case, i.e., for an exponentially small reflection coefficient
one may obtain a Boltzmann-Planck factor from the
$\frac{R_{\omega}}{1-R_{\omega}}$ ratio. The semiclassical reflection
coefficient becomes exponentially small in two cases: either for
the `above barrier' scattering situation or well below the potential
barrier.\cite{str,s} For the massless scalar field in the Schwarzschild
background, the exponentially small
backscattering is the main process because of the low height of the barrier.

{\bf 4}.- I now comment on the close relationship
between the `above barrier' thermal radiation and the well-known Stokes
phenomenon
in the mathematics of asymptotics, as implied by the scattering/WKB
picture.\cite{jwkb}

Stokes phenomenon is a general feature of special functions and mathematical
asymptotics and refers to writing the solution of a partial differential
equation as a linear combination of multivalued functions in some asymptotic
domain of the complex plane. In such cases, which are quite common in physics
and mathematics, the
solution is usually an entire function whereas the components of the linear
combination, being multivalued, will change whenever one is going at least once
around
an arbitrary point of the complex plane, and therefore we have to make use of
a different linear combination. Why then use multivalued functions? A good
answer is to quote R.E. Meyer,\cite{me}, ``{\em the representation by
multivalued functions is the only way in which the wave character can be
displayed with great clarity}''.

The Stokes phenomenon for the thermal scattering effects
in the case of massless scalar $s$-waves refers
to {\em complex} `turning points', i.e., complex roots of the momentum
function $p(r^{*})=\omega ^{2}-V_{s}(r^{*})$,
which enters the eikonal exponent $S$ of the WKB wave functionals
$S(r^{*})=\int dr'^{*} [p(r'^{*})]^{1/2}$.
The locally multivalued functions for the `above barrier' thermal effect
are of the WKB type $u_{\pm}=p^{-1/4}\exp(\pm S(r^{*}))$.
In the asymptotic formalism, one does not take into account the potential
barrier from the propagational point of view. In the original papers of
Hawking, it enters merely through what seems to be reasonable
considerations about the horizon states.\cite{h2} It is well established by
now that the semiclassical approximation breaks down at the
horizon.\cite{th,kv} Briefly, one switches from
eikonal waves to normal modes (roughly speaking,
$p^{-1/4}\rightarrow \omega ^{-1/2}$ and
the eikonal exponent reduces to the spatial exponential factor of a simple
harmonic normal mode, $\exp (\pm i\omega r^{*}$)). Although this appears to be
a consistent and correct procedure at least for massless scalar
fields,\cite{w} eikonal waves, being more general waves (the so-called
quasinormal modes are an important class with an extended literature), lead
to further insight in the problem of black hole radiation and stability.


{\bf 5}.- In conclusion,
the association of thermal radiation with the `above barrier'
scattering may be considered as being different from the usual interpretation
of the Hawking effect in terms of `spontaneous emission'
{\em in all modes} implied by the usual black body interpretation and the
{\em in} - {\em out} formalism.
It is in {\em most of the modes} but not in all of them and is due to the
`above-barrier' scattering.
On the other hand, black holes may show up various types of
WKB tunneling phenomena that can affect the thermality of their
radiance.\cite{pk} Consider the more realistic case of
electromagnetic waves impinging on a small Schwarzschild black hole (we
need a small black hole in order to have a relevant radiation effect).
Then two {\em real} turning points,\cite{f} may be present.
In laboratory physics there exist
well-known thermal emission phenomena - field emission and thermionic
emission of electrons from solid state surfaces,\cite{mod}- which are
well described by the WKB tunneling with two real turning points.

\nonumsection{Acknowledgement}
\noindent
This work was partially supported by the CONACyT Project 458100-5-25844E.


\nonumsection{References}


\end{document}